\documentclass[11pt,onecolumn,amssymb,nofootinbib]{revtex4} 
\usepackage{amsmath, amsthm, amscd, amssymb}
\usepackage{bm}
\usepackage{bbm}
\usepackage{graphicx}
\usepackage{epstopdf}

\usepackage{feyn}
\let\vec\boldvec%

\begin{document}

\title{On spontaneous photon emission in collapse models}

\author{Stephen L. Adler}
\email{adler@ias.edu} \affiliation{Institute for Advanced Study, Einstein Drive, Princeton, NJ 08540, USA}

\author{Angelo Bassi}
\email{bassi@ts.infn.it} \affiliation{Department of Physics, University of
Trieste, Strada Costiera 11, 34151 Trieste, Italy} \affiliation{Istituto
Nazionale di Fisica Nucleare, Trieste Section, Via Valerio 2, 34127 Trieste,
Italy}

\author{Sandro Donadi}
\email{donadi@ts.infn.it} \affiliation{Department of Physics, University of
Trieste, Strada Costiera 11, 34151 Trieste, Italy} \affiliation{Istituto
Nazionale di Fisica Nucleare, Trieste Section, Via Valerio 2, 34127 Trieste,
Italy}

\begin{abstract}
We reanalyze the problem of spontaneous photon emission in collapse models.  We show
that the extra term found by Bassi and D\"urr is present for non-white (colored)
noise, but its coefficient is proportional to the zero frequency Fourier component
of the noise.  This leads one to suspect that the extra term is an artifact.  When
the calculation is repeated with the final electron in a wave packet {\it and} with the
noise confined to a bounded region, the extra term vanishes in the limit of continuum
state normalization.  The result obtained by Fu and by Adler and Ramazano\v glu from application
of the Golden Rule is then recovered.

\end{abstract}
\maketitle

\section{Introduction}

In a previous series of articles~\cite{ref:fu,ref:ar,ref:bd}, the problem of the spontaneous emission of radiation from charged particles, as predicted by collapse models, was analyzed in detail. The interest in this kind of problem arises from the fact that it currently sets the strongest upper bound on these models~\cite{ref:ap,ref:sci}. The analysis of~\cite{ref:fu,ref:ar} has been done by using the CSL model~\cite{ref:csl}; since the equations cannot be solved analytically, a master formula has been derived to the first perturbative order with respect to the collapse parameter $\gamma$.  In~\cite{ref:bd}, the same type of analysis has been done within the QMUPL model~\cite{ref:qmupl}; since this model is mathematically simpler than the CSL model, making only the dipole approximation the equations can be solved explicitly. In this second case ~\cite{ref:bd}, the radiation rate for a free charged particle, when expanded to first order in $\gamma$, turns out to be twice as large as that of~\cite{ref:fu,ref:ar}.

On the other hand, the CSL model for a single particle (or a system of distinguishable particles) reduces to the QMUPL model at the statistical level, in the limit of small distances. Since the radiation formulas do not change under this limit, one would then expect to obtain the same formula whether using the CSL model or the QMUPL model, at the appropriate perturbative order. This discrepancy is the motivation behind this work. As we will discuss in the next Sections, a careful analysis of the origin of this factor of 2 difference, using the CSL model, shows that the difference arises because some matrix elements, which vanish under standard QFT approximations, give instead a finite contribution. While this contribution simply doubles the answer in the free particle case with a white noise, it gives rise to awkward, energy non-conserving terms (not related to the steady increase of particle kinetic energy, a common and well-known feature of collapse models) in the case of a colored noise, leading one to suspect that the extra term is not physical.

In this paper we will discuss all these issues, in order to clarify some mathematical details regarding the derivation of the radiation formula. We will work out in detail the perturbation expansion and re-derive Feynman rules, carefully analyzing all approximations, which are standard in quantum field theory and have been used in~\cite{ref:fu,ref:ar}. As we will see, when one takes into account that the final state of the out-going particle is a wave packet and not a plane wave, and when one also takes into account that the noise is confined to a bounded region, the unexpected contribution coming from the matrix elements  vanishes, eliminating the doubling found in ~\cite{ref:bd}, and confirming the answer of ~\cite{ref:fu,ref:ar}.

\section{The CSL model for charged particles}

In the CSL model, the standard Schr\"odinger equation is modified by adding nonlinear and stochastic
terms which cause the collapse of the wave function. In the It\^o formalism, it takes the following form:
\begin{equation} \label{eq:csl-massa}
d|\psi_t\rangle = \left[-\frac{i}{\hbar}Hdt +
\frac{\sqrt{\gamma}}{m_{0}}\int d\mathbf{x}\, [M(\mathbf{x}) - \langle M(\mathbf{x}) \rangle_t ]
dW_{t}(\mathbf{x}) - \frac{\gamma}{2m_{0}^{2}} \int d\mathbf{x}\,
[M(\mathbf{x}) - \langle M(\mathbf{x}) \rangle_t ]^2 dt\right] |\psi_t\rangle;
\end{equation}
here $H$ is the standard quantum Hamiltonian of the system and the other two
terms induce the collapse. The mass $m_0$ is a reference mass, which is taken
equal to that of a nucleon. The parameter $\gamma$ is a positive coupling
constant which sets the strength of the collapse process, while $M({\bf x})$ is
a smeared mass density operator:
\begin{equation}
M\left(\mathbf{x}\right)=\underset{j}{\sum}m_{j}N_{j}\left(\mathbf{x}\right),\qquad
N_{j}\left(\mathbf{x}\right)=\int
g\left(\mathbf{x-y}\right)
\psi_{j}^{\dagger}\left(\mathbf{y}\right)\psi_{j}\left(\mathbf{y}\right)\, d^3y,
\end{equation}
$\psi_{j}^{\dagger}\left(\mathbf{y},s\right)$,
$\psi_{j}\left(\mathbf{y},s\right)$ being, respectively, the creation and
annihilation operators of a particle of type $j$ at the space point
$\mathbf{y}$. The smearing function $g({\bf x})$ is taken equal to
\begin{equation} \label{eq:nnbnm}
g(\mathbf{x}) \; = \; \frac{1}{\left(\sqrt{2\pi}r_{C}\right)^{3}}\;
e^{-\mathbf{x}^{2}/2r_{C}^{2}},
\end{equation}
where $r_C \sim 10^{-5}\text{cm}$ is the second new phenomenological constant of the model. This is the correlation length of the noise, and Eq.~\eqref{eq:csl-massa} is such that spatial superpositions smaller than $r_C$ are not suppressed, while spatial superpositions larger than $r_C$ are.  $W_{t}\left(\mathbf{x}\right)$ is an
ensemble of independent Wiener processes, one for each point in space, which are responsible for the random character of the evolution; the quantum average $\langle M(\mathbf{x}) \rangle_t = \langle \psi_t | M(\mathbf{x}) | \psi_t \rangle$ is responsible for its nonlinear character.

As shown e.g. in~\cite{ref:fu,ref:im}, the averaged density matrix evolution associated to Eq.~\eqref{eq:csl-massa} can also be derived from a standard Schr\"odinger equation with a random Hamiltonian. Such an equation does not lead to the state vector reduction, because it is linear; nevertheless, since they both reproduce the same noise averaged density matrix evolution, and since physical quantities like the photon emission rate can be computed from the noise averaged density matrix, the non-collapsing equation can equally well be employed to compute such quantities. The advantage of this second approach is that, being based on a linear (stochastic) Schr\"odinger equation, it is much simpler from the computational point of view. In our case, the stochastic Hamiltonian is:
\begin{equation}
H_{\text{\tiny TOT}} = H - \hbar \sqrt{\gamma} \sum_j \frac{m_{j}}{m_{0}}
\int N(\mathbf{y},t)
\psi_{j}^{\dagger}(\mathbf{y}) \psi_{j}(\mathbf{y})\, d^3y
\end{equation}
where:
\begin{equation} \label{eq:sfsoi}
N(\mathbf{y},t) = \int g(\mathbf{y-x})\xi_{t}(\mathbf{x})\, d^3x,
\end{equation}
and $\xi_{t}(\mathbf{x}) = dW_{t}(\mathbf{x})/dt$ is a white noise field, with
correlation function $\mathbb{E}[ \xi_{t}(\mathbf{x}) \xi_{s}(\mathbf{y})] =
\delta(t-s) \delta({\bf x-y})$. As such, $N(\mathbf{x},t)$ is a Gaussian noise
field, with zero mean and correlation function:
\begin{equation} \label{eq:sdfddas}
{\mathbb E}[N(\mathbf{x},t) N(\mathbf{y},s)] \; = \;
\delta(t-s)F({\bf x} - {\bf y}), \qquad  F({\bf x}) \; = \;
\frac{1}{(\sqrt{4 \pi} r_C)^3} e^{-{\bf x}^2/4 r_C^2}.
\end{equation}
The purpose of this article is to reconsider the analysis of the emission of radiation from a
free charged particle, previously discussed in the literature. Accordingly, in the following we will be interested only
in one type of particle, so from now on we will drop the sum over $j$.

The Hamiltonian $H_{\text{\tiny TOT}}$ can be written in terms of an
Hamiltonian density ${\mathcal H}_{\text{\tiny TOT}}$. For the systems we are
interested in studying, we can identify three terms in ${\mathcal
H}_{\text{\tiny TOT}}$:
\begin{equation}
{\mathcal H}_{\text{\tiny TOT}} \; = \; {\mathcal H}_{\text{\tiny P}} +
{\mathcal H}_{\text{\tiny R}} + {\mathcal H}_{\text{\tiny INT}}.
\end{equation}
${\mathcal H}_{\text{\tiny P}}$ contains all terms involving the matter field,
namely its kinetic term, possibly the interaction with an external potential
$V$, and the interaction with the collapsing-noise:
\begin{equation}
{\mathcal H}_{\text{\tiny P}} \; = \; \frac{\hbar^{2}}{2m} \vec{\nabla}\psi^{*}
\cdot \vec{\nabla}\psi \; + \;  V \psi^{*}\psi \; - \; \hbar \sqrt{\gamma} \frac{
m}{m_{0}} N \psi^{*}\psi.
\end{equation}
${\mathcal H}_{\text{\tiny R}}$ contains the kinetic term for the
electromagnetic field:
\begin{equation}
{\mathcal H}_{\text{\tiny R}} = \frac{1}{2}\left(\varepsilon_{0}\mathbf{E}_{\perp}^{2}
+ \frac{\mathbf{B}^{2}}{\mu_{0}}\right),
\end{equation}
where $\mathbf{E}_{\perp}$ is the transverse part of the electric component and
${\bf B}$ is the magnetic component. Finally ${\mathcal H}_{\text{\tiny INT}}$
contains the standard interaction between the quantized electromagnetic field
and the non-relativistic Schr\"odinger field:
\begin{equation}
{\mathcal H}_{\text{\tiny INT}} \; = \; i\frac{\hbar e}{m} \psi^{*}\mathbf{A}\cdot
\vec{\nabla}\psi + \frac{e^{2}}{2m}\mathbf{A}^{2}\psi^{*}\psi.
\end{equation}
The electromagnetic potential ${\bf A}({\bf x},t)$ takes the form:
\begin{equation}
{\bf A}({\bf x},t) \; = \; \sum_{{\bf p}, \lambda} \alpha_p
\left[ \vec{\epsilon}_{{\bf p}, \lambda}\, a_{\bf p} e^{i({\bf p \cdot
x} - \omega_p t)} + \vec{\epsilon}_{{\bf p}, \lambda}^{*}\, a_{\bf p}^{\dagger} e^{-i({\bf p
\cdot x} -
\omega_p t)} \right],
\end{equation}
where $\alpha_{p}=\sqrt{\hbar/2\varepsilon_{0}\omega_{p}L^{3}}$ and
$\omega_{p}=pc$. We are quantizing fields in a cubical box of size $L$. We work in the Coulomb gauge.

To analyze the problem of the emission rate, we will use a perturbative
approach. We identify the unperturbed Hamiltonian as that of the matter field
(interaction with the noise excluded) plus the kinetic term of the
electromagnetic field:
\begin{equation}
\mathcal{H}_{0} \; = \;  \frac{\hbar^{2}}{2m} \vec{\nabla}\psi^{*}\cdot\vec{\nabla}\psi
+ V\psi^{*}\psi + {\mathcal H}_{\text{\tiny R}},
\end{equation}
and we assume that its eigenstates and eigenvalues are known. In particular, we
assume that the matter part $\mathcal{H}_{0}$ is diagonalizable. The perturbed
term then is:
\begin{equation} \label{eq:dsgfdj}
\mathcal{H}_{1} \; = \; i\frac{\hbar e}{m} \psi^{*}\mathbf{A}\cdot\vec{\nabla}\psi
+ \frac{e^{2}}{2m}\mathbf{A}^{2}\psi^{*}\psi - \hbar \sqrt{\gamma} \frac{m}{m_{0}}
N\psi^{*}\psi.
\end{equation}
Such a division of ${\mathcal H}_{\text{\tiny TOT}}$ in $\mathcal{H}_{0} +
\mathcal{H}_{1}$ is justified by the fact that the effects of spontaneous
collapses driven by the noise field are very small at microscopic scales. This
is also true for the electromagnetic effects we are interested in computing.

\section{Feynman rules}

The Feynman diagrams for our model can be derived in a standard way, by means
of the Dyson series and Wick theorem. We will present Feynman rules in space-time, instead of the more familiar Feynman rules in momentum space, because in the following calculation a crucial role will be played by integration over space, and by the large-time limit. They are:

\noindent 1. External lines (the symbol $\bullet$ denotes the generic
space-time vertex $({\bf x}, t)$):

\begin{center}
\includegraphics[width=14cm, keepaspectratio]{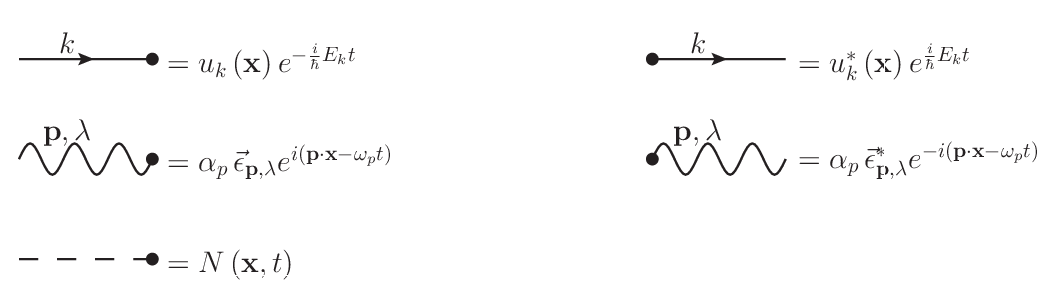}
\end{center}

The functions $u_{k}(\mathbf{x})$ are the eigenstates of
$-\frac{\hbar^{2}}{2m}\vec{\nabla}^{2} + V$, and $E_k$ is the associated
eigenvalue:
\[
\left[-\frac{\hbar^{2}}{2m}\vec{\nabla}^{2} + V\right]u_{k}(\mathbf{x}) \; =
\; E_{k}u_{k}(\mathbf{x}).
\]
Since the noise field $N$ is treated classically, there is no distinction
between incoming and outgoing lines.

\noindent 2. Internal lines. The propagators for the matter field and for the
photons, are:

\begin{center}
{\includegraphics[width=12cm, keepaspectratio]{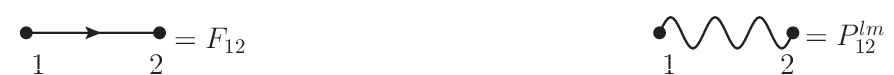}}
\end{center}

with $1 \equiv ({\bf x}_1, t_1)$, $2 \equiv ({\bf x}_2, t_2)$ and:
\begin{eqnarray}
F_{12} & \equiv & F({\bf x}_1,t_1;{\bf x}_2,t_2) \; = \;
\theta\left(t_{2}-t_{1}\right)\underset{k}{\sum}u_{k}\left(\mathbf{x_{2}}\right)
u_{k}^{*}\left(\mathbf{x_{1}}\right)e^{-\frac{i}{\hbar}E_{k}\left(t_{2}-t_{1}\right)}
\label{eq:fgfdssd} \\
P^{lm}_{12} & \equiv & P^{lm}({\bf x}_1,t_1;{\bf x}_2,t_2) \; = \;
\theta\left(t_{1}-t_{2}\right)\underset{\mathbf{k},\lambda}
{\sum}\alpha_{k}^{2}\epsilon_{\mathbf{k},\lambda}^{l}
\epsilon_{\mathbf{k},\lambda}^{*m}e^{i\left[\mathbf{k}\cdot\left(\mathbf{x_{1}}-
\mathbf{x_{2}}\right)-\omega_{k}\left(t_{1}-t_{2}\right)\right]} \nonumber \\
& & \qquad\qquad\qquad\qquad\; + \; \theta\left(t_{2}-t_{1}\right)\underset{\mathbf{k},
\lambda}
{\sum}\alpha_{k}^{2}\epsilon_{\mathbf{k},\lambda}^{m}\epsilon_{\mathbf{k},\lambda}^{*l}
e^{i\left[\mathbf{k}\cdot\left(\mathbf{x_{2}}-\mathbf{x_{1}}\right)-\omega_{k}\left(t_{2}-t_
{1}\right)\right]}.
\end{eqnarray}

\noindent 3. Vertices. There are three types of vertices, corresponding to the
three terms in the interaction Hamiltonian ${\mathcal H}_1$:

\begin{center}
{\includegraphics[width=14cm, keepaspectratio]{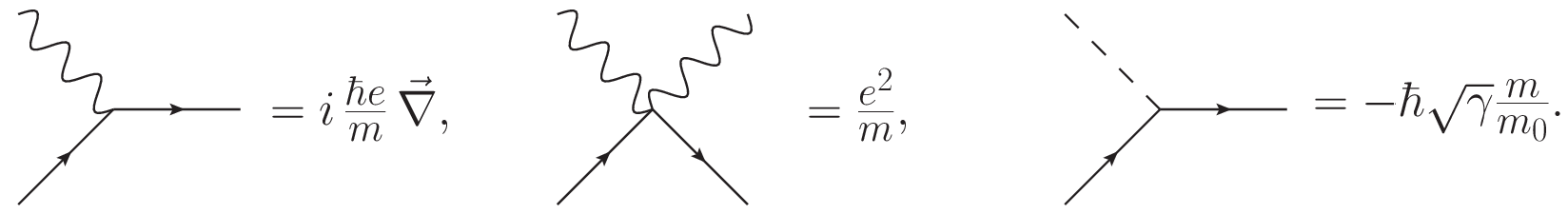}}
\end{center}

In the first vertex, the derivative acts always on the incoming external line.
In the second vertex, $e^2/m$ appears in place of $e^2/2m$ (as one would
naively expect by inspecting at Eq.~\eqref{eq:dsgfdj}) in order to take
properly into account the multiplicity of the diagrams. The same rule applies
also to the standard scalar QED (without the noise term).

\noindent 4. One has to integrate over space and time in all vertices
\[
\frac{1}{ (i\hbar)^n } \prod_{j = 1}^{n} \;
\int_{t_{i}}^{t_{f}}dt_{j}\int_{L^3} d\mathbf{x}_{j}
\]
Note that there is no factorial term $1/n!$ coming from the Dyson's series,
because this is canceled by the multiplicity of the diagram\footnote{More
precisely, a diagram containing $n$ vertices has a factor $1/n!$ in front,
coming from thew Dyson's expansion. However, there are $n!$ such identical
diagrams, differing only in the way the vertices are numbered.}. Only diagrams
containing double photon propagators, like:

\begin{center}
{\includegraphics[width=3cm, keepaspectratio]{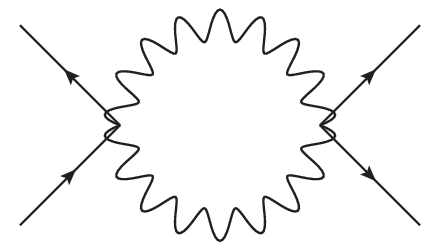}}
\end{center}

do not follow this rule. In such a case, one has to multiply by a factor $1/2$
for each such loop.

\section{Photon emission probability at first perturbative order}

At first order in $\sqrt{\gamma}$ and $e$, the relevant contributions to the process of photon
emission, coming from the interaction of the free particle with the noise
field, are:
\begin{center}
{\includegraphics[width=12cm, keepaspectratio]{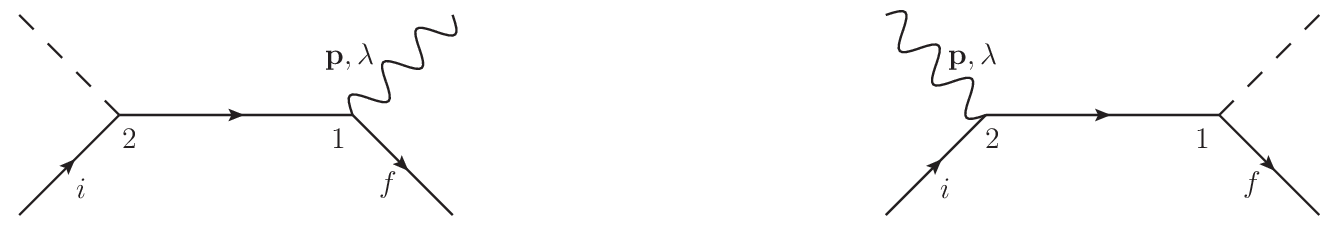}}
\end{center}
According to the rules previously outlined, the contribution of the first
diagram is:
\begin{eqnarray} \label{eq:hetryukeqry}
\lefteqn{-\frac{1}{\hbar^{2}}\,\alpha_{p}\left(i\frac{\hbar
e}{m}\right)\left(-\hbar\sqrt{\gamma}\frac{
m}{m_{0}}\right)\underset{k}{\sum}\int_{t_{i}}^{t_{f}}dt_{1}
\int_{t_{i}}^{t_{1}}dt_{2}\;
e^{i\omega_{p}t_{1}}e^{-\frac{i}{\hbar}E_{i}t_{2}}
e^{\frac{i}{\hbar}E_{f}t_{1}}e^{-\frac{i}{\hbar}E_{k}\left(t_{1}-t_{2}\right)}}
\qquad\qquad\qquad\qquad\nonumber \\
& &
\times \int_{L^3} d\mathbf{x}_{1} \int_{L^3} d\mathbf{x}_{2}\, u_{i}(\mathbf{x}_{2}) e^{-i
\mathbf{p}
\cdot\mathbf{x}_{1}} \vec{\epsilon}_{\mathbf{p},\lambda}^{*} \cdot [\vec{\nabla}u_{k}(\mathbf
{x}_{1})]
u_{k}^{*}\left(\mathbf{x}_{2}\right)
  u_{f}^{*}(\mathbf{x}_{1}) N({\bf x}_2, t_2),
\qquad\qquad
\end{eqnarray}
while the contribution of the second diagram is:
\begin{eqnarray}
\lefteqn{-\frac{1}{\hbar^{2}}\,\alpha_{p}\left(i\frac{\hbar
e}{m}\right)\left(-\hbar\sqrt{\gamma}\frac{
m}{m_{0}}\right)\underset{k}{\sum}\int_{t_{i}}^{t_{f}}dt_{1}
\int_{t_{i}}^{t_{1}}dt_{2}\;
e^{i\omega_{p}t_{2}}e^{-\frac{i}{\hbar}E_{i}t_{2}}
e^{\frac{i}{\hbar}E_{f}t_{1}}e^{-\frac{i}{\hbar}E_{k}\left(t_{1}-t_{2}\right)}}
\qquad\qquad\qquad\qquad\nonumber \\
& &
\times \int_{L^3} d\mathbf{x_{1}} \int_{L^3} d\mathbf{x}_{2}\; u_{k}(\mathbf{x}_{1}) u_{k}^
{*}(\mathbf{x}_{2}) e^{-i\mathbf{p}
\cdot\mathbf{x}_{2}} \vec{\epsilon}_{\mathbf{p},\lambda}^{*} \cdot
[\vec{\nabla} u_{i}(\mathbf{x}_{2})] u_{f}^{*}(\mathbf{x}_{1}) N({\bf x}_1, t_1).
\qquad\qquad
\end{eqnarray}
Summing these two contributions, the transition amplitude $T_{fi}$ becomes:
\begin{eqnarray} \label{eq:sdajfdst}
T_{fi} & = & -\frac{1}{\hbar^{2}}\alpha_{p}\left(i\frac{\hbar e}{m}\right)
\left(-\hbar \sqrt{\gamma}\frac{m}{m_{0}}\right)\underset{k}{\sum}\int_{t_{i}}^{t_{f}}dt_{1}
\int_{t_{i}}^{t_{1}}dt_{2}\; e^{\frac{i}{\hbar}\left(E_{f}-E_{k}\right)
t_{1}}e^{\frac{i}{\hbar}\left(E_{k}-E_{i}\right)t_{2}} \nonumber \\
& \times & \left[\left\langle f\right|e^{i\omega_{p}t_{1}}e^{-i\mathbf{p}
\cdot\hat{\mathbf{x}}}\vec{\epsilon}_{\mathbf{p},\lambda}^{*}\cdot\vec{\nabla}\left|k\right
\rangle
\left\langle k\right|N(\hat{{\bf x}}, t_{2})
\left|i\right\rangle
 +
\left\langle f\right|N(\hat{{\bf x}}, t_{1})
\left|k\right\rangle \left\langle k\right|e^{i\omega_{p}t_{2}}
e^{-i\mathbf{p}\cdot\hat{\mathbf{x}}}\vec{\epsilon}_{\mathbf{p},\lambda}^{*}
\cdot \vec{\nabla}\left|i\right\rangle
\right],\;\;\;\;\;
\end{eqnarray}
where we have introduced the position operator $\hat{{\bf x}}$. It is
convenient to rewrite the above expression in a more compact form. Since the
correlation function~\eqref{eq:sdfddas} of the noise is a product of its
time and space components, as far as the average values are concerned we
can replace $N({\bf x}, t)$ with $\xi_t \, N({\bf x})$, where $\xi_t$ is a
white noise in time, while $N({\bf x})$ is a Gaussian noise in space, with zero
mean and correlator $F({\bf x} - {\bf y})$. We also introduce the following two
operators:
\begin{equation}
{\mathcal{R}}^{p} \; \equiv \; \alpha_{p}
\left(i\frac{\hbar e}{m}\right) e^{-i\mathbf{p}\cdot\mathbf{x}}
\vec{\epsilon}_{\mathbf{p},\lambda}\cdot \vec{\nabla}, \qquad\quad
{\mathcal{N}} \; \equiv \;
-\hbar \sqrt{\gamma}\frac{m}{m_{0}}\; N(\hat{{\bf x}}).
\end{equation}
The first operator refers to the radiative contribution (hence the symbol
${\mathcal{R}}$), the second one to the interaction with the noise (hence the
symbol ${\mathcal{N}}$). Defining moreover the matrix elements ${\mathcal
R}_{ki}^{p} \equiv \langle k | {\mathcal R}^{p} |i \rangle $ and ${\mathcal
N}_{ki} \equiv \langle k | {\mathcal N} |i \rangle $, considering photons
with linear polarization ($\vec{\epsilon}_{\mathbf{p},\lambda}^{*} =
\vec{\epsilon}_{\mathbf{p},\lambda}$), we can write Eq.~\eqref{eq:sdajfdst} in
the following way:
\begin{eqnarray} \label{eq:ta_fin}
\lefteqn{T_{fi} \; = \;
-\frac{1}{\hbar^{2}}\underset{k}{\sum}\int_{t_{i}}^{t_{f}}dt_{1}
\int_{t_{i}}^{t_{1}}dt_{2}\,
e^{\frac{i}{\hbar}(E_{f}-E_{k})t_{1}}
e^{\frac{i}{\hbar}t(E_{k}-E_{i})t_{2}}} \qquad\qquad\qquad\qquad && \nonumber \\
& & \qquad\qquad\times \left[e^{i\omega_{p}t_{1}}\xi_{t_{2}}
{\mathcal R}_{fk}^{p}{\mathcal N}_{ki}
+
e^{i\omega_{p}t_{2}}\xi_{t_{1}}
{\mathcal N}_{fk}{\mathcal R}_{ki}^{p}
\right].
\end{eqnarray}
This is the final expression of the first-order transition amplitude for a
charged particle to emit a photon, as a consequence of the interaction with the
noise field. The particle might be free, as we will consider in the next
section, or interacting with an external potential.

\section{Emission rate for a free particle}
\label{sec:two}

In the case of a free charged particle, the initial and final states and the
generic eigenstate of $H_{P}$ are:
\begin{equation}
u_{i}(\mathbf{x}) = \frac{1}{\sqrt{L^{3}}}, \qquad\quad
u_{f}(\mathbf{x}) = \frac{e^{i\mathbf{q}\cdot\mathbf{x}}}{\sqrt{L^{3}}}, \qquad\quad
u_{k}(\mathbf{x}) = \frac{e^{i\mathbf{k}\cdot\mathbf{x}}}{\sqrt{L^{3}}},
\end{equation}
and we have chosen the reference frame where the particle is initially at rest.
The corresponding eigenvalues are $E_{k}= \hbar^{2}\mathbf{k}^{2}/2m$, and
similarly for $E_i$ and $E_f$. The matrix elements for the radiative part can
now be easily computed:
\begin{eqnarray}
{\mathcal R}_{fk}^{p} & = & \langle f |{\mathcal R}^{p} | k \rangle
= \frac{1}{L^{3}} \int_{L^3} d\mathbf{x}\,
e^{-i\mathbf{q}\cdot\mathbf{x}} \left[\alpha_{p}\left(i\frac{\hbar
e}{m}\right) e^{-i\mathbf{p}\cdot\mathbf{x}}
\vec{\epsilon}_{\mathbf{p},\lambda} \cdot \vec{\nabla}\right]
e^{i\mathbf{k}\cdot\mathbf{x}}
\nonumber \\
& & \qquad\qquad\, = \alpha_{p}\left(-\frac{\hbar e}{m}\right)\left(\vec{\epsilon}_{\mathbf
{p},\lambda}
\cdot \mathbf{q}\right)\delta_{\mathbf{k,q+p}}, \label{eq:ppo} \\
{\mathcal N}_{ki} & = & -\hbar \sqrt{\lambda} \frac{m}{m_0} \frac{1}{L^3} \int d {\bf x} \, N({\bf x}) e^{-i {\bf k} \cdot {\bf x}} \label{eq:1bis }\\
{\mathcal R}_{ki}^{p} & = & \langle k |{\mathcal R}^{p} | i \rangle = 0, \label{eq:1}\\
{\mathcal N}_{fk} & = & -\hbar \sqrt{\lambda} \frac{m}{m_0} \frac{1}{L^3} \int d {\bf x} \, N({\bf x}) e^{i ({\bf k} - {\bf q}) \cdot {\bf x}}  \label{eq:1ter}
\end{eqnarray}
As we see, the contribution given by the second Feynman diagram is null.
Therefore, in squaring Eq.~\eqref{eq:ta_fin}, taking the average with respect
to the noise, we obtain the relatively simple expression:
\begin{eqnarray} \label{eq:dfgdff}
\lefteqn{\mathbb{E} |T_{fi}|^{2} = \frac{1}{\hbar^{4}}\underset{k}{\sum}\underset{j}{\sum}
\; {\mathcal R}_{fj}^{p*} {\mathcal R}_{fk}^{p} {\mathbb E}
[{\mathcal N}_{ji}^{*} {\mathcal N}_{ki}] }
\qquad\qquad\qquad & & \\
& & \times \int_{0}^{t}dt_{1}\int_{0}^{t_{1}}dt_{2}\int_{0}^{t}dt_{3}\int_{0}^{t_{3}}dt_{4}
e^{iat_{1}}e^{ibt_{2}}e^{ict_{3}}e^{idt_{4}}\delta\left(t_{2}-t_{4}\right),
\nonumber
\end{eqnarray}
where we have set $t_i = 0$ and $t_f = t$, and moreover we have defined the
constants:
\begin{equation} \label{eq:con1}
a\equiv\frac{1}{\hbar}\left(E_{f}+\hbar\omega_{p}-E_{k}\right),\;\;\;
b\equiv\frac{1}{\hbar}\left(E_{k}-E_{i}\right),\;\;\;
c\equiv-\frac{1}{\hbar}\left(E_{f}+\hbar\omega_{p}-E_{j}\right),\;\;\;
d\equiv-\frac{1}{\hbar}\left(E_{j}-E_{i}\right),
\end{equation}
We focus the attention on the temporal part. We have:
\begin{eqnarray}
T & = & \int_{0}^{t}dt_{1}\int_{0}^{t_{1}}dt_{2}\int_{0}^{t}dt_{3}\int_{0}^{t_{3}}dt_{4}
e^{iat_{1}}e^{ibt_{2}}e^{ict_{3}}e^{idt_{4}}\delta\left(t_{2}-t_{4}\right) \nonumber \\
& = & \int_{0}^{t}dt_{1}\int_{0}^{t}dt_{2}\int_{0}^{t}dt_{3}\int_{0}^{t}dt_{4}
e^{iat_{1}}e^{ibt_{2}}e^{ict_{3}}e^{idt_{4}}\delta\left(t_{2}-t_{4}\right)
\theta\left(t_{1}-t_{2}\right)\theta\left(t_{3}-t_{4}\right) \nonumber \\
& = & \int_{0}^{t}dt_{2}\int_{t_{2}}^{t}dt_{1}\int_{t_{2}}^{t}dt_{3}e^{iat_{1}}e^{i\left(b+d
\right)t_{2}}e^{ict_{3}} \nonumber \\
& = & \frac{1}{ca}\left[e^{i\left(c+a\right)t}\frac{1-e^{igt}}{ig}+e^{iat}
\frac{e^{i\left(g+c\right)t}-1}{i\left(g+c\right)}+e^{ict}
\frac{e^{i\left(g+a\right)t}-1}{i\left(g+a\right)}+
\frac{1-e^{i\left(g+c+a\right)t}}{i\left(g+c+a\right)}\right], \label{eq:integrale
tempi metà}
\end{eqnarray}
where we have defined:
\begin{equation} \label{eq:con2}
g \; \equiv \; b+d \; = \; \frac{1}{\hbar}\left(E_{k}-E_{j}\right).
\end{equation}
Because of the relation $a+b+c+d=a+c+g=0$, Eq.~\eqref{eq:integrale tempi metà}
simplifies to:
\begin{equation} \label{eq:gfgg}
T \; = \; \frac{1}{ac}\left[\frac{e^{-igt}-1}{ig}+\frac{e^{iat}-1}{ia}+\frac{e^{ict}-1}{ic}-t
\right],
\end{equation}

We are now ready to replace the matrix elements ${\mathcal R}_{fj}^{p*}$ and
${\mathcal R}_{fk}^{p}$ with the explicit expressions~\eqref{eq:ppo} for the
free particle. The indices $k, j$ become vector indices ${\bf k,j}$ labeling
the wave number, and the constraints given by the deltas in the ${\mathcal R}$
terms (see Eq.~\eqref{eq:ppo}) suppress the two sums in Eq.~\eqref{eq:dfgdff}.
This moreover implies:
\begin{eqnarray}
g & = & 0 \\
a & = & -c \; = \; \frac{1}{\hbar}\left(E_{f}+\hbar\omega_{p}-E_{q+p}\right) \; = \;
\left(pc-\frac{\hbar\mathbf{p}^{2}}{2m}-\frac{\hbar\mathbf{q\cdot p}}{m}\right), \label
{eq:ffou}
\end{eqnarray}
Then expression~\eqref{eq:gfgg} for $T$ simplifies to:
\begin{equation}
T \; = \; \frac{2}{a^{3}}\left[at-\sin\left(at\right)\right],
\end{equation}


We now focus on the remaining part of Eq.~\eqref{eq:dfgdff}:
\begin{equation}
\frac{1}{\hbar^{4}}\underset{k}{\sum}\underset{j}{\sum} \; {\mathcal
R}_{fj}^{p*} {\mathcal R}_{fk}^{p} {\mathbb E} [{\mathcal
N}_{ji}^{*} {\mathcal N}_{ki}] \; = \;
\frac{1}{\hbar^{4}}\alpha_{p}^{2}\left(\frac{\hbar
e}{m}\right)^{2}\left(\vec{\epsilon}_{\mathbf{p},\lambda}\cdot\mathbf{q}\right)^{2}
{\mathbb E} [{\mathcal N}_{\left(p+q\right)i}^{*} {\mathcal
N}_{\left(p+q\right)i}],
\end{equation}
where we have taken into account the constraints coming from the Kronecker delta in Eq.~\eqref
{eq:ppo}. The stochastic average gives:
\begin{equation} \label{eq:aaasspdft}
{\mathbb E} [{\mathcal N}_{\left(p+q\right)i}^{*} {\mathcal
N}_{\left(p+q\right)i}] \; = \;
\hbar^{2}\gamma\left(\frac{m}{m_0}\right)^{2}\frac{1}{L^6}\int_{L^3}d\mathbf{x_1}\int_{L^3}d
\mathbf{x_2}\,
e^{i\mathbf{\left(p+q\right)\cdot\left(x_1-x_2\right)}}F\left(\mathbf{x_1-x_2}\right).
\end{equation}
We make the change of variable: $\mathbf{x=x_1-x_2}$ and
$\mathbf{y=x_1+x_2}$ (the Jacobian is $1/8$) and we use the
rule:
\begin{equation}
\int_{-\frac{L}{2}}^{+\frac{L}{2}}dx_1\int_{-\frac{L}{2}}^{+\frac{L}{2}}dx_2f\left
(x_1,x_2\right)=
\frac{1}{2}\int_{0}^{+L}dx\int_{-\left(L-x\right)}^{+\left(L-x\right)}dy\left[f\left(x,y
\right)+f\left(-x,y\right)\right],
\end{equation}
thus obtaining:
\begin{eqnarray}
\lefteqn{{\mathbb E} [{\mathcal N}_{\left(p+q\right)i}^{*} {\mathcal
N}_{\left(p+q\right)i}] \; = \;} \nonumber \\
& = & \hbar^{2}\gamma\left(\frac{m}{m_0}\right)^{2}\frac{1}{8 L^6}
\prod_{i=1}^{3} \int_{0}^{+L}d{x_i}
\int_{-\left(L-x_i\right)}^{+\left(L-x_i\right)}d{y_i}\frac{1}{\sqrt{4\pi}r_c}
\left[e^{i(p+q)_ix_i}+e^{-i(p+q)_ix_i}\right]e^{-x_i^2/4r_C^2}.
\end{eqnarray}
The integral over $y_i$ gives:
\begin{equation}
\frac{1}{2L}\int_{-\left(L-x_i\right)}^{+\left(L-x_i\right)}d{y_i} \; = \; 1-\frac{x_i}{L}.
\end{equation}
The second term vanishes in the large $L$ limit, so we can ignore it. We are left with:
\begin{equation}
{\mathbb E} [{\mathcal N}_{\left(p+q\right)i}^{*} {\mathcal
N}_{\left(p+q\right)i}] \; = \;\hbar^{2}\gamma\left(\frac{m}{m_0}\right)^{2}\frac{1}{L^3}
\int_{-L}^{+L}d\mathbf{x}e^{i\mathbf{\left(p+q\right)\cdot
x}}F\left(\mathbf{x}\right).
\end{equation}
In the large $L$ limit, the integral gives the Fourier transform of the correlation function $F$. Taking into account the form~\eqref{eq:sdfddas} of $F$, and collecting all pieces, we have:
\begin{equation}\label{eq:result1}
\mathbb{E} |T_{fi} |^{2} \; = \; \Lambda
(\vec{\epsilon}_{\mathbf{p},\lambda} \cdot
\mathbf{q})^{2}e^{-\mathbf{\left(q+p\right)}^{2}r_{C}^{2}}\;\;
\frac{at - \sin(at)}{a^3},
\end{equation}
where:
\begin{equation}\label{eq:gamma}
\Lambda \; = \;
2\, \frac{1}{\hbar^{4}}\,\alpha_{p}^{2}\left(\frac{\hbar
e}{m}\right)^{2} \hbar^{2} \gamma \left(\frac{m}{m_{0}}\right)^{2} \frac{1}{L^{3}} \; = \; \frac{1}{L^6} \frac{\gamma\hbar
e^{2}}{\varepsilon_{0}cm_{0}^{2}p}
\end{equation}
collects all constant terms.

The emission rate $\Gamma({\bf p})$ can be computed from the transition
probability, by differentiating over time, and by summing over the momentum
${\bf q}$ of the outgoing particle and the polarization $\lambda$ of the
emitted photon, according to the formula:
\begin{equation}
\frac{d\Gamma}{d^{3}p} \; = \; \left(\frac{L}{2\pi}\right)^{6}\int d\mathbf{q}
\; \underset{\lambda}{\sum}\; \frac{\partial}{\partial
t}\; \mathbb{E}| T_{fi} |^{2}.
\end{equation}
Let us choose the axes so that ${\bf p} = (0,0,p)$; in this way $a$, as given
by Eq.~\eqref{eq:ffou}, becomes a function only of $p$ and $q_z$.  The sum over
polarizations then gives $\sum_{\lambda} (\vec{\epsilon}_{\mathbf{p},\lambda}
\cdot \mathbf{q})^{2} = q_x^2 + q_y^2$. All factors $L$ cancel with each other,
so we can take safely the limit $L \rightarrow + \infty$. The sum over ${\bf
q}$ then becomes a triple integral. The two integrals over $q_x$ and $q_y$ can
be easily computed, being Gaussian, and we obtain:
\begin{equation} \label{eq:rate}
\frac{d\Gamma}{d^{3}p} \; = \; 2 \Lambda \left(
\frac{\sqrt{\pi}}{r_C}\right) \left( \frac{\sqrt{\pi}}{2 r_C^3}
\right) \int dq_z\, e^{-(q_z + p)^2 r_C^2}\;\;
\frac{1-\cos(at)}{a^{2}}.
\end{equation}
The above integral can be rewritten in the following way:
\begin{equation}
\int dq_z\, e^{-(q_z + p)^2 r_C^2}  \frac{1-\cos(at)}{a^{2}} \; = \;
\frac{m}{\hbar p} \int dz \, e^{-z^2\beta^2}\;\; \frac{1 - \cos[(D -
z)t]}{(D - z)^2},
\end{equation}
where we have defined the following new quantities: $z = \hbar p (q_z + p)/m$,
$D = pc + \hbar p^2/2m$ and $\beta = m r_c/ \hbar p$. Since $\beta \simeq
10^{-13}\textrm{ s}$ and $D \simeq pc \simeq 10^{19}\textrm{ s}^{-1}$ for a non
relativistic electron and for radiation in the KeV region, the Gaussian term in
Eq.~\eqref{eq:rate} is vanishing small in the region where $1 - \cos[(D - z)t]/(D - z)^2$ is most appreciably different from zero. Around the origin, where the Gaussian is not negligible, the denominator varies slowly, and one can approximate $1/(D - z)^2 \sim 1/D^2$, and bring it out of the integral. What remains, apart the Gaussian term, is $1 - \cos[(D - z)t]$. The second term oscillates very rapidly and gives a negligible contribution to the integral. Thus only the first term survives.
When integrating moreover over all directions in which the photon can be
emitted, the emission rate becomes:
\begin{equation} \label{eq:rate quasi finale}
\frac{d\Gamma}{dp} \; = \; \frac{\lambda\hbar e^{2}}{2\pi^{2}\varepsilon_{0}c^{3}m_{0}^{2}r_
{c}^{2}p},
\end{equation}
with $\lambda \equiv \gamma/8\pi^{3/2}r_{C}^{3}$ equal to the collapse rate
first introduced in the GRW model~\cite{ref:grw}. In the above expression, we have
neglected the oscillating term, which averages to zero over typical
experimental times. The above result is expressed in SI units. The
transformation to CGS units simply requires the replacement $\varepsilon_{0}
\rightarrow 1/4\pi$, in which case we obtain twice the results of~\cite{ref:fu,ref:ar}.

The mathematical reason for such a difference lies in the type of
approximations used to obtain the final formula.  Going back to
Eq.~\eqref{eq:gfgg}, in~\cite{ref:ar} the following approximation was made:
\begin{equation}
\frac{1}{ac}\left[\frac{e^{-igt}-1}{ig}+\frac{e^{iat}-1}{ia}+
\frac{e^{ict}-1}{ic}-t\right]\simeq-\frac{t}{ac}.
\end{equation}
While this is legitimate in general, it gives problems in the free particle
case. Here, as we have seen, $g=0$, meaning that the oscillating term depending
on $g$ becomes linear in $t$. This contribution sums with the other linear
term, giving the factor of 2 difference. We also note that the remaining
two oscillating terms are mathematically important, though physically
negligible. Since for a free particle $a=-c$, they reduce to a cosine, which
makes sure that the integral in Eq.~\eqref{eq:rate} is convergent. Without it,
the pole at the denominator would produce a divergence.

\section{Emission rate in the non-white noise case}

To better understand the origin of the factor of 2 difference in the white-noise case, we generalize now to
the colored noise case, where we will find that the extra term which doubles the answer of ~\cite{ref:fu,ref:ar}
has a suspicious energy non-conserving form\footnote{Moreover, an unpublished calculation by S.L.A. shows that for an electron bound in a hydrogen atom,
the extra term leads to a suspicious orders of magnitude increase in the radiation rate, rather than just the
doubling found in the free particle case.}, which is not related to the steady increase of particle kinetic
energy due to energy transfer from the noise to the particle during the collapse, a well-known feature of these models.
To see this, we now generalize Eq.~\eqref{eq:rate quasi finale} to the case where the collapsing
noise has a correlation function which is not white in time:
\begin{equation} \label{eq:sdfddasbis2}
{\mathbb E}[N(\mathbf{x},t) N(\mathbf{y},s)] \; = \;
f(t-s)F({\bf x} - {\bf y}).
\end{equation}

We can start from
Eq.~\eqref{eq:dfgdff} for the average transition probability:
\begin{eqnarray} \label{eq:dfgdffbis}
\lefteqn{\mathbb{E} |T_{fi}|^{2} =
\frac{1}{\hbar^{4}}\underset{k}{\sum}\underset{j}{\sum} \int_{L^3}
d\mathbf{z}\; {\mathcal R}_{fk}^{p} {\mathcal N}_{ki}(\mathbf{z})
{\mathcal R}_{fj}^{p*} {\mathcal N}_{ji}^{*}(\mathbf{z}) }
\qquad\qquad\qquad & & \\
& & \times \int_{0}^{t}dt_{1}\int_{0}^{t_{1}}dt_{2}\int_{0}^{t}dt_{3}\int_{0}^{t_{3}}dt_{4}
e^{iat_{1}}e^{ibt_{2}}e^{ict_{3}}e^{idt_{4}}f\left(t_{2}-t_{4}\right),
\nonumber
\end{eqnarray}
where now $f$ replaces the Dirac delta. The coefficients $a,b,c$ and $d$ are
the same as in~\eqref{eq:con1}. The only effect of the non-white noise is to
modify the time dependent part of the transition probability, which we consider
separately:
\begin{equation}
T \; = \; \int_{0}^{t}dt_{1}\int_{0}^{t_1}dt_{2}\int_{0}^{t}dt_{3}\int_{0}^{t_3}dt_{4}\,
e^{iat_{1}}e^{ibt_{2}}e^{ict_{3}}e^{idt_{4}}
f(t_{2}-t_{4}).
\end{equation}
This can be rewritten as follows:
\begin{eqnarray}
T & = & \int_{0}^{t}dt_{2}\int_{t_{2}}^{t}dt_{1}\int_{0}^{t}dt_{4}\int_{t_{4}}^{t}
dt_{3}\,e^{iat_{1}}e^{ibt_{2}}e^{ict_{3}}e^{idt_{4}}f\left(t_{2}-t_{4}\right) \nonumber \\
& = & -\frac{1}{ac}\int_{0}^{t}dt_{2}\int_{0}^{t}dt_{4}\left(e^{iat}-e^{iat_{2}}\right)
\left(e^{ict}-e^{ict_{4}}\right)e^{ibt_{2}}e^{idt_{4}}f\left(t_{2}-t_{4}\right).
\label{eq:op8pa}
\end{eqnarray}
There are four terms, which all have the following structure:
\begin{equation}
I \equiv \int_{0}^{t}du \int_{0}^{t}dv\, e^{i\alpha
u} e^{i\beta v} F(u-v) = \int_{0}^{t}du \int_{0}^{t}dv\, e^{\frac{i}{2}
\left[\left(\alpha+\beta\right)\left(u+v\right)+\left(\alpha-\beta\right)
\left(u-v\right)\right]}f\left(u-v\right).
\end{equation}
We perform the change of variable $x = u-v$ and $y = u+v$. In these new
variables, the integral changes as follows:
\begin{equation}
\int_{0}^{t}du \int_{0}^{t}dv\, f(u,v) \; = \; \frac{1}{2}\int_{0}^{t}dx
\int_{x}^{2t-x}dy\, \left[f(-x,y) + f(x,y)\right].
\end{equation}
In our case, the integrating variables separate, and the integral over $y$ can
be easily performed, giving:
\begin{equation}
\int_{x}^{2t-x}dy\, e^{\frac{i}{2}\left(\alpha+\beta\right)y} \; = \;
4 \, \frac{e^{\frac{i}{2}\left(\alpha+\beta\right)t}}{(\alpha+\beta)}\,
\sin{\frac{1}{2}(\alpha+\beta)(t-x)}.
\end{equation}
Then, taking into account that $f(u-v) = f(x) = f(-x)$, the double integral $I$
reduces to:
\begin{equation}
I \; = \; 4\, \frac{e^{\frac{i}{2}(\alpha+\beta)t}}{(\alpha+\beta)}
\int_{0}^{t}dx\, f(x) \sin\frac{1}{2}(\alpha+\beta)(t-x) \cos\frac{1}{2}(\alpha-\beta)x.
\end{equation}

Going back to Eq.~\eqref{eq:op8pa}, taking into account the relation $a+b+c+d =
0$, we can write:
\begin{eqnarray} \label{eq:hdfgzds}
T & = & -\frac{4}{ac} \left\{ \frac{e^{-\frac{i}{2}g t}}{g}
\int_{0}^{t}dx\, f(x)\sin\left[\frac{1}{2}g(t-x)\right] \cos\left[\frac{1}{2}(b-d)x\right]
\right. \nonumber \\
& & \qquad\;\;\, -
\frac{e^{\frac{i}{2}at}}{a} \int_{0}^{t}dx\, f(x) \sin\left[\frac{1}{2}a(t-x)\right]
\cos\left[\frac{1}{2}ax\right] \nonumber \\
& & \qquad\;\;\, -
\frac{e^{\frac{i}{2}ct}}{c} \int_{0}^{t}dx\, f(x) \sin\left[\frac{1}{2}c(t-x)\right]
\cos\left[\frac{1}{2}cx\right] \nonumber \\
& & \left. \qquad\quad\;\;\, +
\frac{1}{2} \int_{0}^{t}dx\, f(x)(t-x)\cos\left[(a+b)x\right]\right\}.
\end{eqnarray}
Of course, in the white noise case $f(x) = \delta(x)$, the above
equation reduces to Eq.~\eqref{eq:gfgg}.\footnote{Note that
$\int_0^t dx \delta(x)g(x)=\frac{1}{2}g(0)$, for a general function
$g(x)$, must be used
 in the reduction to the white noise case.} In computing the matrix elements ${\mathcal
R}_{ij}^{p}$ and ${\mathcal N}_{ij}$ for the free particle, a further
constraint comes from the Kronecker delta of Eq.~\eqref{eq:ppo}, which implies
$a = - c$ and $g = 0$. Accordingly, the expression for $T$ further simplifies
to:
\begin{eqnarray}
T & = & \frac{2}{a^{2}} \left\{ \int_{0}^{t}dx\, f(x)(t-x)\left[ \cos(bx)+
\cos[(a+b)x] \right] \right. \nonumber \\
& - & \left. \frac{4}{a}\cos\left( \frac{1}{2}at\right) \int_{0}^{t}dx\, f(x)
\sin\left[\frac{1}{2}a(t-x)\right] \cos\left(\frac{1}{2}ax\right) \right\}.
\end{eqnarray}
The next step, in computing the emission rate, is to compute the time
derivative. Differentiating with respect to the upper limit of the integrals,
produces terms proportional to $f(t)$, which vanish in the large time limit, as
we assume that the correlation function has a finite correlation time. The
remaining terms coming from the second line produce oscillating terms, which
average to zero. Thus, the only significant term, in the large time limit, is:
\begin{equation}
\frac{\partial T}{\partial t} \; \xrightarrow[t \,\rightsquigarrow \, \infty]{} \;
\frac{1}{a^2} \left[ \tilde{f}(b) + \tilde{f}(a + b) \right],
\end{equation}
where we have defined the Fourier transform of the correlation function:
\begin{equation} \label{eq:gdskteyfdah}
\tilde{f}(\omega) \; \equiv \; 2\int_{0}^{+\infty} dt\, f(t) \cos(\omega t) \;
= \; \int_{-\infty}^{+\infty} dt\, f(t) e^{i \omega t}.
\end{equation}
Finally, in computing the integral over the final momentum of the particle (see
Eq.~\eqref{eq:rate}), we have approximated the Gaussian term by a Dirac
delta, meaning that we are imposing $\mathbf{q} \simeq - \mathbf{p}$. This
implies:
\begin{eqnarray}
a & = & \left( pc - \frac{\hbar \mathbf{p}^2}{2m} - \frac{\hbar \mathbf{p} \cdot
\mathbf{q}}{m} \right) \; \longrightarrow \left(pc + \frac{\hbar \mathbf{p}^2}{2m}
\right) \; \simeq \; pc \label{eq:gsddsfsd}\\
b & = & \frac{\hbar(\mathbf{q+p})^{2}}{2m} \; \longrightarrow \; 0.
\end{eqnarray}
Thus we have:
\begin{equation}
\left. \frac{d \Gamma}{dp} \right|_{\text{\tiny NON-WHITE}} \; = \; \frac{1}{2}
\left[ \tilde{f}(0) + \tilde{f}(pc) \right]
\times \left. \frac{d \Gamma}{dp} \right|_{\text{\tiny WHITE}}.
\end{equation}
The second term is the expected one: the probability of emitting a photon with momentum $p$ is proportional to the weight of the Fourier component of the noise corresponding to the frequency $\omega_p = pc$. The first term instead is independent of the photon's momentum, and looks suspicious. Precisely this term, in the white-noise limit, is responsible for the factor of 2 difference, as one can easily check. In the remaining sections we analyze the origin of such an unexpected term.

\section{Computation using a generic final state for the charged particle}

At the end of Section~\ref{sec:two} we have discussed that the factor of 2 difference arises because $g$ as defined in Eq.~\eqref{eq:con2} becomes 0, the reason being that the deltas in the ${\mathcal R}$ terms (see Eq.~\eqref{eq:ppo}) force $E_k$ to be equal to $E_j$. One could then expect that by considering a generic final state for the outgoing particle---in place of the more artificial plane wave---such a constraint is removed. We discuss such a possibility in this Section\footnote{In appendix B we review the formalism for computing the transition amplitude to a final wave packet state.}. So, instead of a final state for the particle with definite momentum, let us now take a normalized wave packet:

\begin{equation}
u_{i}\left(\mathbf{x}\right)=\frac{1}{\sqrt{L^{3}}},\qquad
u_{f}\left(\mathbf{x}\right)=\underset{\bf \Delta}{\sum}h\left({\bf \Delta}\right)
\frac{e^{i\left(\mathbf{q+\Delta}\right)\cdot\mathbf{x}}}{\sqrt{L^{3}}},\qquad
u_{k}\left(\mathbf{x}\right)=\frac{e^{i\mathbf{k}\cdot\mathbf{x}}}{\sqrt{L^{3}}},
\end{equation}
where $h\left(\Delta\right)$ normalizes the wave function:
\begin{equation}
1=\int_{L^3} d\mathbf{x}\left|\psi_{f}\right|^{2}=\frac{1}{L^{3}}
\underset{{\bf \Delta}'}{\sum}\underset{\bf \Delta}{\sum}h^{*}
\left({\bf \Delta}'\right)h\left({\bf \Delta}\right)\int_{L^3} d\mathbf{x}
e^{-i\left(\mathbf{q+\Delta'}\right)\cdot\mathbf{x}}
e^{i\left(\mathbf{q+\Delta}\right)\cdot\mathbf{x}}=
\underset{\bf \Delta}{\sum}\left|h\left({\bf \Delta}\right)\right|^{2}.
\end{equation}
The matrix elements~\eqref{eq:ppo}--\eqref{eq:1ter} now become:
\begin{eqnarray}
{\mathcal R}_{fk}^{p} & = & \left\langle f\left|{\mathcal R}^{p}\right|k\right\rangle =
\underset{\bf \Delta}{\sum}h^{*}\left({\bf \Delta}\right)\left\langle
f_{\bf \Delta}\left|{\mathcal R}^{p}\right|k\right\rangle
=\alpha_{p}\left(-\frac{e\hbar}{m}\right)\underset{\bf \Delta}
{\sum}h^{*}\left({\bf \Delta}\right)\left[\mathbf{\epsilon_{p}}\cdot
\left(\mathbf{q+\Delta}\right)\right]\delta_{\mathbf{k,q+\Delta+p}}.\quad \label{eq:fdgdfg} \\
{\mathcal N}_{ki} & = & -\hbar \sqrt{\lambda} \frac{m}{m_0} \frac{1}{L^3} \int d {\bf x} \, N({\bf x}) e^{-i {\bf k}\cdot {\bf x}} \\
{\mathcal R}_{ki}^{p} & = & \left\langle k\left|{\mathcal R}^{p}\right|i\right\rangle \; = \;
0, \\
{\mathcal N}_{fk} & = & -\hbar \sqrt{\lambda} \frac{m}{m_0} \frac{1}{L^3} \underset{\bf \Delta}{\sum}h\left({\bf \Delta}\right) \int d {\bf x} \, N({\bf x}) e^{i ({\bf k} - {\bf q} - {\bf \Delta}) \cdot {\bf x}}  \label{eq:1ter2}
\end{eqnarray}
Since also in this case ${\mathcal R}_{ki}^{p}=0$, the formula of
$\mathbb{E}|T_{fi}|^{2}$ is still given by Eq.~\eqref{eq:dfgdff}, where the temporal part takes the same form as in~\eqref{eq:gfgg}:
\begin{equation} \label{eq:mod2b}
\mathbb{E}|T_{fi}|^{2} \; = \;
\frac{1}{\hbar^{4}}\underset{k}{\sum}\underset{j}{\sum}\, {\mathcal R}_{fj}^{p*} {\mathcal R}_{fk}^{p}\, {\mathbb E}[{\mathcal N}_{ji}^{*} {\mathcal N}_{ki}] \; T,
\end{equation}
with $T$ given in Eq.~\eqref{eq:gfgg}.
The two Kronecker deltas coming from Eq.~\eqref{eq:fdgdfg} set: $\mathbf{k=q+\Delta+p}$ and  $\mathbf{j=q+\Delta'+p}$.
Accordingly, the coefficients $a$, $c$ and $g$, defined in~\eqref{eq:con1}
and~\eqref{eq:con2}, become:
\begin{equation} \label{eq:fgf}
a=\frac{\left(E_{f}+\hbar\omega_{p}-E_{q+\Delta+p}\right)}{\hbar},\qquad
c=-\frac{\left(E_{f}+\hbar\omega_{p}-E_{q+\Delta'+p}\right)}{\hbar},\qquad
g=\frac{E_{q+\Delta+p}-E_{q+\Delta'+p}}{\hbar}.
\end{equation}
Moreover, we have:
\begin{equation} \label{eq:cgxcgdfs}
{\mathbb E}[{\mathcal N}_{ji}^{*} {\mathcal N}_{ki}] \; = \; \hbar^2 \gamma \left( \frac{m}{m_0} \right)^2 \frac{1}{L^6} \int_{L^3} d {\bf x}_1 \int_{L^3} d {\bf x}_2 e^{-i({\bf p+q})\cdot ({\bf x}_1 - {\bf x}_2)} e^{-i ({\bf \Delta}\cdot {\bf x}_1 - {\bf \Delta}'\cdot {\bf x}_2)} F({\bf x}_1 - {\bf x}_2).
\end{equation}
The two exponents can be rewritten as: $-i({\bf p+q})\cdot ({\bf x}_1 - {\bf x}_2) - (i/2)({\bf \Delta} - {\bf \Delta}') \cdot ({\bf x}_1 - {\bf x}_2) - (i/2)({\bf \Delta} + {\bf \Delta}') \cdot ({\bf x}_1 + {\bf x}_2)$. We now make the change of variables: ${\bf x} = {\bf x}_1 - {\bf x}_2$, ${\bf y} = {\bf x}_1 + {\bf x}_2$, as we did after Eq.~\eqref{eq:aaasspdft}. The integral over ${\bf y}$ produces $L^{3}\delta_{\mathbf{\Delta,\Delta'}}$ plus extra terms which vanish in the large $L$ limit. Thus, as in the previous section, $a$,
$c$ and $g$ take the values:
\begin{equation}
a \; = \; -c \; = \; \frac{\left(E_{f}+\hbar\omega_{p}-E_{q+\Delta+p}\right)}{\hbar},
\qquad g \; = \; 0.
\end{equation}
The integral over ${\bf x}$ gives the Fourier transform of the correlator $F$; accordingly, the transition probability reduces to:
\begin{equation} \label{eq:gfgffg}
\mathbb{E}|T_{fi}|^{2} \; = \; \Lambda
\underset{\Delta}{\sum}\left|h\left(\Delta\right)\right|^{2}
[\mathbf{\epsilon_{p}}\cdot\left(\mathbf{q+\Delta}\right)]^{2}
e^{-\mathbf{\left(\mathbf{q+\Delta+p}\right)}^{2}r_{C}^{2}}
\left[ \frac{at - \sin (at)}{a^{3}} \right],
\end{equation}
with $\Lambda$ defined as in~\eqref{eq:gamma}. As we see, the structure is minimally modified
from that of Eq. \eqref{eq:result1}, and the answer of [3] for the reduction rate is still obtained.

\section{Computation with a noise confined in space}

The calculation of the previous Section shows that the reason why $g=0$ also for an outgoing wave packet, is because a $\delta_{\mathbf{\Delta,\Delta'}}$ appears, which arises from the integral over space with respect to the variable ${\bf y} = {\bf x}_1 + {\bf x}_2$. This suggests that the problem can be avoided by considering a noise which is confined to a finite region of space. We analyze this case here.

Let us suppose that the correlation function of the noise is:
\begin{equation} \label{eq:sdfddas1}
{\mathbb E}[N(\mathbf{x},t) N(\mathbf{y},s)] \; = \;
f(t-s)F({\bf x} - {\bf y}) e^{-({\bf x} + {\bf y})^2/\ell^2},
\end{equation}
where $\ell$ is an appropriate cut off. We start from Eq.~\eqref{eq:mod2b}
\begin{equation} \label{eq:mod3n}
\mathbb{E}|T_{fi}|^{2} \; = \;
\frac{1}{\hbar^{4}}\underset{k}{\sum}\underset{j}{\sum}\, {\mathcal R}_{fj}^{p*} {\mathcal R}_{fk}^{p}\, {\mathbb E}[{\mathcal N}_{ji}^{*} {\mathcal N}_{ki}]
\,T,
\end{equation}
where the temporal part $T$ is given by Eq.~\eqref{eq:hdfgzds}. The two Kronecker deltas coming from Eq.~\eqref{eq:fdgdfg} set: $\mathbf{k=q+\Delta+p}$ and $\mathbf{j=q+\Delta'+p}$. In this case, Eq.~\eqref{eq:cgxcgdfs} is replaced by:
\begin{equation}
{\mathbb E}[{\mathcal N}_{ji}^{*} {\mathcal N}_{ki}] \; = \; \hbar^2 \gamma \left( \frac{m}{m_0} \right)^2 \frac{1}{L^6} \int_{L^3} d {\bf x}_1 \int_{L^3} d {\bf x}_2 e^{-i({\bf p+q})\cdot ({\bf x}_1 - {\bf x}_2)} e^{-i ({\bf \Delta}\cdot {\bf x}_1 - {\bf \Delta}'\cdot {\bf x}_2)} F({\bf x}_1 - {\bf x}_2) e^{-({\bf x} + {\bf y})^2/\ell^2}.
\end{equation}
As before, we perform the change of variables: $\bf{x} = {\bf x}_{1} - {\bf x}_{2}$, $\bf{y} = {\bf x}_{1} + {\bf x}_{2}$. In integrating over the new variables, we use the rule:
\begin{equation}
\int_{-\frac{L}{2}}^{+\frac{L}{2}}dx_{1}\int_{-\frac{L}{2}}^{+\frac{L}{2}}dx_{2}f\left(x_{1},x_{2}\right)
=\frac{1}{2}\int_{0}^{L}dx\int_{-\left(L-x\right)}^{+\left(L-x\right)}dy
\left[f\left(x,y\right)+f\left(-x,y\right)\right].
\end{equation}
In our case:
\begin{eqnarray}
f(\mathbf{x_{1},x_{2}}) & = & e^{i\mathbf{(q+\Delta'+p)\cdot x_{1}}}e^{-i\mathbf{(q+\Delta+p)\cdot x_{2}}}F(\mathbf{x_{1}-x_{2}})e^{-(\mathbf{x_{1}+x_{2}})^{2}/l^{2}} \\
& = & e^{i\left({\bf q}+\frac{{\bf \Delta}'+{\bf \Delta}}{2}+ {\bf p}\right)\cdot\mathbf{x}}e^{\frac{i}{2}\mathbf{(\Delta'-\Delta)\cdot\mathbf{y}}}F(\mathbf{x})
e^{-\mathbf{y}^{2}/l^{2}}=\prod_{i=1}^3 f_i(x_i,y_i),\\
f_i(x_i,y_i)& =
&e^{\frac{i}{2}(j+k)_ix_i}e^{\frac{i}{2}(\Delta'-\Delta)_i
y_i}\frac{1}{\sqrt{4\pi}r_C} e^{-x_i^2/4r_C^2}e^{-y_i^2/l^2},
\end{eqnarray}
where we have used the Kronecker delta constraints to replace
$\mathbf{q+p}+\frac{1}{2}\left(\mathbf{\Delta+\Delta'}\right)$ by
$\frac{1}{2}\left(\mathbf{j+k}\right)$. Thus we arrive at the
following expression:
\begin{eqnarray}
\mathbb{E}[\mathcal{N}_{ji}^{*}\mathcal{N}_{ki}] &=&
\hbar^{2}\gamma\left(\frac{m}{m_{0}}\right)^{2}\frac{1}{8L^{6}}\prod_{i=1}^{3}\int_{0}^{L}\!\!\! dx_{i}\int_{-\left(L-x_{i}\right)}^{+\left(L-x_{i}\right)}\!\!\!\!\!\!\!\! dy_{i}\; 2\cos\!\left[\frac{1}{2}(j+k)_i x_i \right]\nonumber \\
&\times&
e^{\frac{i}{2}(\Delta'-\Delta)_iy_i}\frac{1}{\sqrt{4\pi}r_C}e^{-x_i^2/4r_C^2}e^{-y_i^2/l^2}.
\nonumber \\
& &
\end{eqnarray}
Since $e^{-x_i^2/4r_C^2}$ has a cutoff at $|x_i| \sim r_C \ll L$,
$x_i$ never approaches $L$. So we can write (in the large $L$
limit):
\begin{eqnarray}\label{eq:expression3}
\mathbb{E}[\mathcal{N}_{ji}^{*}\mathcal{N}_{ki}] & = & \hbar^{2}\gamma\left(\frac{m}{m_{0}}\right)^{2}\frac{1}{L^{3}}\prod_{i=1}^{3}\int_{0}^{L}dx_{i}\; 2\cos\left[\frac{1}{2}(j+k)_i x_i \right] \frac{1}{\sqrt{4\pi}r_C}e^{-x_i^2/4r_C^2} \nonumber \\
& \times & \prod_{j=1}^{3}\left(\frac{1}{2L}\int_{-L}^{+L}dy_{j}\right)
e^{\frac{i}{2}(\Delta'-\Delta)_j y_j} e^{-y_j^2/l^{2}}.
\end{eqnarray}

To summarize, substituting Eq. \eqref{eq:expression3} into Eq. \eqref{eq:mod3n} and noting the Kronecker delta in ${\mathcal R}_{fk}^{p}$,  the effect of having considered a final wave packet in place of a plane wave, and of having confined the noise in space, is that the double Kronecker delta $\delta_{\mathbf{k,q+p}} \delta_{\mathbf{j,q+p}}$ is replaced by:
\begin{equation} \label{eq:k1}
K_{jk} \; = \; \sum_{\bf \Delta} \sum_{\bf \Delta'} h^{*}({\bf \Delta}) h({\bf \Delta}') \delta_{\mathbf{k,q+\Delta+p}} \delta_{\mathbf{j,q+\Delta'+p}}\prod_{j=1}^{3}\left(\frac{1}{2L}\int_{-L}^{+L}dy_{j}\right)e^{\frac{i}{2}\mathbf{(\Delta'-\Delta)\cdot\mathbf{y}}}e^{-\mathbf{y}^{2}/l^{2}}.
\end{equation}
One can easily check that when $\ell = \infty$, the triple integral reduces to $\delta_{{\bf \Delta}, {\bf \Delta}'}$. $K_{jk}$ then becomes: $\sum_{\bf \Delta} | h({\bf \Delta})|^2 \delta_{\mathbf{k,q+\Delta+p}} \delta_{\mathbf{j,q+\Delta+p}}$, which implies ${\bf j} = {\bf k}$. The same happens when $\ell < \infty$, but $h({\bf \Delta}) = \delta_{{\bf \Delta}, {\bf 0}}$, i.e. when the final state is a plane wave. Thus both a final wave-packet state and a noise confined in space are necessary in order to avoid the factor of 2 term.

Coming back to Eq.~\eqref{eq:mod3n}, we have:
\begin{eqnarray}
\mathbb{E}|T_{fi}|^{2} & = & \gamma \alpha_p^2 \left( \frac{e}{m_0} \right)^2 \frac{1}{(2\pi)^6} \int d\, {\bf k} \int d\, {\bf j}\; h({\bf j} - {\bf q} - {\bf p}) h^{*}({\bf k} - {\bf q} - {\bf p}) (\mathbf{\epsilon_{p}}\cdot \mathbf{j}) (\mathbf{\epsilon_{p}}\cdot \mathbf{k}) \nonumber \\
& & \prod_{i=1}^{3}\int_{0}^{\infty}dx_{i}\; 2\cos\left[\frac{1}{2}(j+k)_i x_i \right] \frac{1}{\sqrt{4\pi}r_C}e^{-x_i^2/4r_C^2} \frac{1}{8} \prod_{j=1}^{3}\int_{-\infty}^{+\infty}dy_{j} e^{\frac{i}{2}(j-k)_j y_j} e^{-y_j^2/l^{2}} T, \qquad
\end{eqnarray}
where we have used the expression~\eqref{eq:fdgdfg} for ${\mathcal R}_{fk}^{p}$ (and we simplified the formula using the Kronecker delta) and we have performed the large $L$ limit. We can now compute both the integrals over $x_i$ and over $y_j$:
\begin{eqnarray}
\mathbb{E}|T_{fi}|^{2} & = & \gamma \alpha_p^2 \left( \frac{e}{m_0} \right)^2 \frac{1}{(2\pi)^6} \int d\, {\bf k} \int d\, {\bf j}\; h({\bf j} - {\bf q} - {\bf p}) h^{*}({\bf k} - {\bf q} - {\bf p}) (\mathbf{\epsilon_{p}}\cdot \mathbf{j}) (\mathbf{\epsilon_{p}}\cdot \mathbf{k}) \nonumber \\
& & e^{-({\bf j} + {\bf k})^2 r_C^2/4} \left( \frac{\sqrt{\pi} \ell}{2} \right)^3 e^{- ({\bf j} - {\bf k})^2 \ell^2 / 16}\;T, \qquad
\end{eqnarray}
Since we can take $\ell$ arbitrarily large, according to the second Gaussian term, only those elements with ${\bf j} \simeq {\bf k}$ are relevant. We can therefore simplify the above expression as follows:
\begin{equation}
\mathbb{E}|T_{fi}|^{2} \; = \; \gamma \alpha_p^2 \left( \frac{e}{m_0} \right)^2 \frac{1}{(2\pi)^6} \int d\, {\bf k} |h({\bf k} - {\bf q} - {\bf p})|^2  (\mathbf{\epsilon_{p}}\cdot \mathbf{k})^2 e^{-{\bf k}^2 r_C^2} \left( \frac{\sqrt{\pi} \ell}{2} \right)^3 \int d\, {\bf j}\;  e^{- ({\bf j} - {\bf k})^2 \ell^2 / 16}\;T. \qquad
\end{equation}
Here we have used the wave packet assumption that $h$ is a smooth function of its arguments, and not a delta function.  

We now focus the attention on the last integral containing the time dependence, which generates the factor of 2 problem when $\ell = \infty$. We now show that the undesired term has a vanishing contribution, for large times. For simplicity, we focus our attention to the white-noise expression for $T$ (see Eq.~\eqref{eq:gfgg}), but the calculation can immediately be generalized also to the non-white-noise case of Eq.~\eqref{eq:hdfgzds} as well. Taking into consideration only the undesired term, the integral to compute becomes:
\begin{equation}
J = \left( \frac{\sqrt{\pi} \ell}{2} \right)^3 \int d {\bf j} \, e^{- ({\bf j} - {\bf k})^2 \ell^2 / 16}\; \frac{1}{ac} \frac{e^{-igt} - 1}{ig}.
\end{equation}
According to Eq.~\eqref{eq:con1}, when ${\bf j} = {\bf k}$, then $c = -a$, and the dependence of $c$ over ${\bf j}$ drops out. According to Eq.~\eqref{eq:con2}, $g = \hbar ({\bf k}^2 - {\bf j}^2)/2m \simeq \hbar {\bf k} \cdot ({\bf k} - {\bf j})/m$.  Moreover we can re-write the exponential term in integral form. We arrive at the following formula:
\begin{eqnarray}
J & = & \left( \frac{\sqrt{\pi} \ell}{2} \right)^3 \frac{1}{a^2} \int_{0}^{t} ds \int d {\bf j}\; e^{- ({\bf j} - {\bf k})^2 \ell^2 / 16 -i\hbar {\bf k} \cdot ({\bf k} - {\bf j})s/m} \nonumber \\
& = &
\frac{(2\pi)^3}{a^2} \int_{0}^{t} ds e^{- 4 \hbar^2 {\bf k}^2 s^2 / m^2 \ell^2}.
\end{eqnarray}
As we see, if we take the limit $\ell \rightarrow \infty$, the integral here above gives a linear increase in time, and therefore contributes to the total rate, giving rise to the factor of 2 problem. On the other hand, if we keep $\ell$ finite, we compute the rate (which corresponds to differentiating in time) and we  take the large time limit, such a term decays exponentially and does not contribute to the asymptotic rate. 
All the conditions here above are consistent with typical experimental situations.  In this regime,
the extra term found in [3] is negligible, and the Golden Rule formula used as the basis for the
calculations of [1], [2] gives the entire answer.

We pose the
question: If the calculation is repeated without a spatial cutoff on
the noise, but with the initial and final electron wave functions
taking the interaction with the noise into account (in analogy with
the distorted wave Born approximation), will the extra term found in
[3] then be suppressed?

\section{An alternative, simpler calculation}

We give here an alternative calculation, that proceeds from the transition amplitude {\it before} squaring and
averaging over the noise. Focusing only on the time-dependent part of the transition amplitude $T_{fi}$, we have\footnote{See Eq.~\eqref{eq:ta_fin}, with ${\mathcal R}_{ki}^{p} = 0$. For simplicity, we take $t_i = 0$ and $t_f = t$. The coefficients $a$ and $b$ are defined in Eq.~\eqref{eq:con1}.}:
\begin{align}\label{eq:tamplitude}
T_{fi} &\propto \int_0^{t}dt_1 \int_0^{t_1} dt_2 \exp{(iat_1)} \exp{(ibt_2)} \xi_{t_2}\cr
=& \int_0^t dt_2 \int_{t_2}^t dt_1  \exp{(iat_1)} \exp{(ibt_2)} \xi_{t_2}~~~,\cr
\end{align}
with $\xi_{t_2}$ the temporal part of the noise.
Let us now
follow the treatment of first order time-dependent perturbation theory given in the text of Schiff \cite{ref:schiff}, and
assume that the noise acts only during the finite time interval between $t=0$ and $t=t_0$.  So we take for
$\xi_{t_2}$ the form
\begin{equation}\label{eq:noisefourier}
\xi_{t_2}=\theta(t_2)\theta(t_0-t_2) \int d\omega  N(\omega) \exp{i\omega t_2} ~~~,
\end{equation}
where we have expressed the noise amplitude in terms of its frequency Fourier transform.
Substituting Eq.~\eqref{eq:noisefourier} into Eq. \eqref{eq:tamplitude} we get
\begin{align}\label{eq:tcalculation}
T_{fi} \propto& \int d\omega N(\omega) \int_0^{t_0} dt_2 \int_{t_2}^t dt_1  \exp{(iat_1)} \exp{[i(\omega+b)t_2]} \cr
=& \int d\omega N(\omega) \phi(\omega,a,b)~~~,\cr
\end{align}
with $\phi(\omega,a,b)$ given by
\begin{equation}\label{eq:phidef}
\phi(\omega,a,b)= -\exp{(iat)} \frac{\exp{[i(\omega+b) t_0]}-1}{a(\omega +b)}
+\frac{\exp{[i(a+b+\omega) t_0]} -1}{a(a+b+\omega)}~~~.
\end{equation}
Near $\omega=-(a+b)$, the function $\phi(\omega,a,b)$ is dominated by the second term,
which has squared modulus
\begin{equation}\label{eq:secondterm}
\frac{1}{a^2} \frac{\sin^2{[\frac{1}{2}(a+b+\omega)t_0]} } {[\frac{1}{2}(a+b+\omega)]^2}
\end{equation}
which for large $t_0$ is effectively
\begin{equation}\label{eq:largetime1}
\frac{2\pi t_0}{a^2} \delta(\omega+a+b)~~~.
\end{equation}
Similarly, near $\omega=-b$, the function $\phi(\omega,a,b)$ is dominated by the first term,
which has squared modulus
\begin{equation}\label{eq:secondterm}
\frac{1}{a^2} \frac{\sin^2{[\frac{1}{2}(b+\omega)t_0]} } {[\frac{1}{2}(b+\omega)]^2}
\end{equation}
which for large $t_0$ is effectively
\begin{equation}\label{eq:largetime2}
\frac{2\pi t_0}{a^2} \delta(b+\omega)~~~.
\end{equation}
So if we take the absolute value squared of $T_{fi}$ and average over the noise amplitude
$N(\omega)$ using $ \mathbb{E}[N(\omega)N(\omega^{\prime})]=\delta(\omega-\omega^{\prime})\tilde{f}(\omega)$,
with $\tilde{f}(\omega)$ the power spectral function of the noise introduced in Eq.~\eqref{eq:gdskteyfdah}, the second term contributes
the energy conserving term found previously, while the first term contributes the extra energy
non-conserving term. However, as already noted, physical measurements detect wave packet
final states, not plane wave energy eigenstate final states.  The amplitude $T_{fi}$ for transition
to a wave packet final state is obtained by averaging with a weighting function over a small interval
of final state energies, peaked around the observed mean values. This is equivalent to replacing $\phi(\omega,a,b)$ by an average
over $a$ with a weight $w(a)$ (which for convenience we take to have unit integral $\int da w(a) =1$)
\begin{equation}\label{eq:averagephi}
\int da w(a) \phi(\omega,a,b)= -\Big[\int da w(a)\exp{(iat)}\Big] \frac{\exp{[i(\omega+\bar{b}) t_0]}-1}{\bar{a}(\omega +\bar{b})}
+ \frac{\exp{[i(\bar{a}+\bar{b}+\omega) t_0]} -1}{\bar{a}(\bar{a}+\bar{b}+\omega)}~~~,
\end{equation}
where $\bar{a} \simeq pc$ is the expected value of $a$ (see Eq.~\eqref{eq:gsddsfsd}) and $\bar{b}$ the expected value of $b$.
The second term is unchanged by this averaging, but the first term now contains a factor
\begin{equation}\label{eq:averagefirst}
\Big[\int da w(a)\exp{(iat)}\Big]~~~,
\end{equation}
which approaches zero by the Riemann-Lebesgue lemma as $t\to \infty$ for fixed $t_0$.  So the
energy non-conserving term drops out of the large time amplitude for transition to a wave packet
state, and only the expected energy conserving term remains.  Note that the two ingredients of the
calculation here, (i) use of a finite time cutoff for the noise, and (ii) use of wave packet
final states, correspond to the ingredients used in the calculation of the preceding section, where
we assumed a spatially  bounded noise, and spatial wave packet final states.

\section{Spontaneous emission from the vacuum}

The Feynman rules suggest that also the following process is possible:
\begin{center}
{\includegraphics[width=5cm, keepaspectratio]{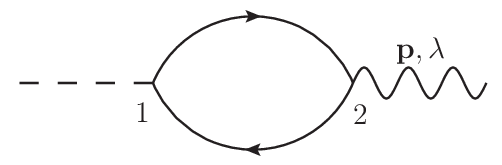}}
\end{center}

\noindent which corresponds to a photon emitted from the vacuum. The analytical
expression for such a process contains two internal particle's propagators
$F_{12} F_{21} \varpropto \theta(t_2 - t_1) \theta(t_1 - t_2)$ (see
Eq.~\eqref{eq:fgfdssd}), giving a zero contribution. Thus at the
non-relativistic level there is no spontaneous photon emission from the vacuum.
However, such a process is expected not to vanish at the relativistic level. We leave this computation for future research.

\section*{Acknowledgements}

A.B. and S.D. acknowledge support from NANOQUESTFIT, the COST Action MP1006 and INFN, Italy. A.B. wishes to acknowledge the hospitality of the Institute for Advanced Study in Princeton, where part of this work has been done. S.L.A. acknowledges the hospitality of the Abdus Salam International Centre for Theoretical
Physics, where this work was completed.  He also acknowledges support of the Department of Energy
under grant DE-FG02-90ER40642.

\section*{Appendix A: Noise in the box}

Let us consider a Gaussian noise, with zero mean and correlation function:
\begin{equation}
{\mathbb E}[N(\mathbf{x},t), N(\mathbf{y},s)] \; = \; \delta(t-s)
F(\mathbf{x-y}).
\end{equation}
Writing it in Fourier components:
\begin{equation}
N(\mathbf{x},t) \; = \; \frac{1}{(2\pi)^{3}}\int
d\mathbf{k}\, e^{i\mathbf{k\cdot x}}\tilde{N}(\mathbf{k},t),
\end{equation}
we easily find the following relation for the correlation function in momentum
space:
\begin{equation} \label{eq:gdfgoo}
{\mathbb E}[\tilde{N}(\mathbf{k},t), \tilde{N}(\mathbf{k'},s)] \; = \;
(2\pi)^{3} \delta(t-s) \delta(\mathbf{k+k'}) \tilde{F}(\mathbf{k}),
\end{equation}
where:
\begin{equation}
\tilde{F}(\mathbf{k}) \; \equiv \; \int d{\bf x}\, e^{-i {\bf k \cdot x}} F({\bf x})
\end{equation}
is the Fourier transform of the spatial correlator. In placing the noise in a
box of size $L$, we select only the Fourier components with the correct
boundary conditions. therefore we define:
\begin{equation} \label{eq:xzzdf}
N_{L}(\mathbf{x},t) \; \equiv \; \frac{1}{L^{3}}\sum_{\mathbf{j}=-\infty}^{+\infty}e^{i\frac
{2\pi}{L}\mathbf{j\cdot
x}}\tilde{N}_{L}\left(\mathbf{j},t\right), \qquad
\tilde{N}_{L}\left(\mathbf{j},t\right)\equiv\tilde{N}\left(\frac{2\pi}{L}\mathbf{j},t\right).
\end{equation}
From Eq.~\eqref{eq:gdfgoo} on can write the correlation function of
$\tilde{N}_{L}\left(\mathbf{j},t\right)$:
\begin{equation}
{\mathbb E}[\tilde{N}_{L}(\mathbf{j},t),
\tilde{N}_{L}(\mathbf{j'},s)] \; = \;
L^{3}\delta(t-s)\delta_{\mathbf{j,-j'}}\tilde{F}(\frac{2\pi}{L}\mathbf{j}),
\end{equation}
from which one finds the following correlator for the noise in the box
$N_{L}(\mathbf{x},t)$:
\begin{equation}
{\mathbb E}[N_{L}(\mathbf{x},t), N_{L}(\mathbf{y},s)]
\; = \; \delta(t-s) F_{L}(\mathbf{x-y'}),
\end{equation}
with:
\begin{equation} \label{eq:xcxcxv}
F_{L}(\mathbf{x-y}) \; \equiv \;
\frac{1}{L^{3}}\sum_{\mathbf{j}=-\infty}^{+\infty}e^{i\frac{2\pi}{L}\mathbf{j\cdot
(x-x')}}\tilde{F}(\frac{2\pi}{L}\mathbf{j}).
\end{equation}
One can easily prove that in the limit $L \rightarrow \infty$, the noise
$N_{L}(\mathbf{x},t)$ as defined in Eq.~\eqref{eq:xzzdf}, converges to
$N(\mathbf{x},t)$, and the correlation function $F_{L}(\mathbf{x-y})$ as
defined in Eq.~\eqref{eq:xcxcxv}, converges to $F(\mathbf{x-y})$.

\section*{Appendix B: Transition Amplitude to a Wave Packet}

We review here how to calculate the transition amplitude to a wave packet final state.
We start from:
\begin{align}\label{eq:gentime}
|\psi_t\rangle  =& U(t,t_0) |\psi_0\rangle  \cr
=& \exp{(-i{\cal H}_0 t/\hbar)} U_I(t,t_0) \exp{(i{\cal H}_0 t/\hbar)} |\psi_0\rangle~~~,  \cr
\end{align}
with ${\cal H}_0$ the unperturbed Hamiltonian and $U_I(t,t_0)$ the interaction picture
time evolution operator.  Let us take the initial state $|i\rangle \equiv|\psi_0\rangle$ to be a single
electron at rest, as in the text, so that ${\cal H}_0 |\psi_0\rangle =0$. Multiplying on the left by a
complete set of energy eigenstates $1=\sum_n |E_n\rangle \langle E_n|$ (with the energy here including
both final electron and photon energies) we get
\begin{align}\label{eq:gentime1}
|\psi_t\rangle  =&\sum_n \langle E_n| U(t,t_0) |i\rangle | E_n \rangle\cr
=&\sum_n \langle E_n|  U_I(t,t_0)|i \rangle \exp{(-i E_n t/\hbar)}|E_n \rangle   \cr
=&\sum_n T_{ni} \exp{(-i E_n t/\hbar)}|E_n \rangle ~~~,  \cr
\end{align}
with $T_{ni}=\langle E_n|  U_I(t,t_0)|i \rangle$ the transition amplitude as defined in the text.  The modulus squared $|T_{ni}|^2$
of this amplitude gives the probability of finding an outgoing energy
eigenstate  solution $\exp{(-i E_n t/\hbar)}|E_n \rangle$ of the unperturbed time dependent Schr\"odinger equation.
However, in realistic experiments, we measure not exact energy eigenstates, but instead wave packets
that are superpositions of energy eigenstates over a narrow energy range.  We can form a complete set of
such wave packets by summing energy eigenstates with a suitable complete set of orthonormal weighting coefficients
$\langle E_n| P\rangle$, with $P$ the parameter set describing the wave packets,  giving for the wave packet basis
\begin{equation}\label{eq:wavebasis}
|P,t\rangle = \sum_n  \langle E_n| P\rangle  \exp{(-i E_n t/\hbar)}|E_n \rangle ~~~.
\end{equation}
If we take the weighting coefficients $ \langle E_n| P\rangle $ to be time-independent, the members of this basis
will also be solutions of the unperturbed time dependent Schr\"odinger equation. The inverse
transformation from wave packet basis to energy eigenstate is given by
\begin{equation}\label{eq:inversebasis}
 \exp{(-i E_n t/\hbar)}|E_n \rangle=\sum_P   \langle P| E_n\rangle |P,t\rangle ~~~.
\end{equation}
Substituting Eq. \eqref{eq:inversebasis} into Eq. \eqref{eq:gentime1} we get
\begin{equation}\label{eq:gentime2}
|\psi_t\rangle = \sum_P \left(\sum_n T_{ni} \langle P| E_n\rangle \right)|P,t\rangle ~~~,
\end{equation}
showing that the transition amplitude to the wave packet state $|P,t\rangle $
is the weighted sum over energy eigenstates $\sum_n T_{ni} \langle P| E_n\rangle $.
When the wave packet also has finite spatial extent, the analogous transition amplitude
to the wave packet state will be a weighted sum of $T_{ni}$ over final energies and momenta.
Equation \eqref{eq:gentime2}, and its generalization to spatial wave packets, justifies the procedure for calculating the transition amplitude to wave
packets used in the text.


\begin{thebibliography}{99}

\bibitem{ref:fu} Q. Fu, {\it Phys. Rev. A}Ê{\bf 56}, 1806 (1997).

\bibitem{ref:ar} S.L. Adler and F.M. Ramazano\v glu, {\it J. Phys. A} {\bf 40}, 13395 (2007).

\bibitem{ref:bd} A. Bassi and D. D\"urr, {\it J. Phys. A} {\bf 42}, 485302 (2009).

\bibitem{ref:ap} S.L. Adler, {\it J. Phys. A} {\bf 40}, 2935 (2007).

\bibitem{ref:sci} S.L. Adler and A. Bassi, {\it Science} {\bf 325}, 275 (2009).

\bibitem{ref:csl} G.C. Ghirardi, P. Pearle and A. Rimini, {\it Phys. Rev. A} {\bf 42}, 78 (1990).

\bibitem{ref:qmupl} L. Di\'osi, {\it Phys. Rev. A} {\bf 40}, 1165 (1989); {\bf 42}, 5086 (1990).

\bibitem{ref:im} S.L. Adler and A. Bassi, {\it J. Phys. A} {\bf 40}, 15083 (2007). See also references therein.

\bibitem{ref:grw} G.C. Ghirardi, A. Rimini and T. Weber, {\bf Phys. Rev. D} {\bf 34}, 470 (1986).

\bibitem{ref:schiff} L. I. Schiff, {\it Quantum Mechanics, Third Edition}, Mc-Graw Hill, New York(1968),
pp. 282-283.

\end{thebibliography}
\end{document}